\documentclass[iop,apj]{emulateapj}

\newcommand{\kms}          {\mbox{${\rm km~s^{-1}}$}}
\def\simgt{\lower.5ex\hbox{$\; \buildrel > \over \sim \;$}}
\def\simlt{\lower.5ex\hbox{$\; \buildrel < \over \sim \;$}}

\slugcomment{Accepted for publication in ApJ}

\shorttitle{Misaligned binary disk system in Orion}
\shortauthors{Williams et al.}

\begin{document}
\title{ALMA observations of a misaligned binary protoplanetary disk system in Orion}

\author{
Jonathan P. Williams\altaffilmark{1},
Rita K. Mann\altaffilmark{2},
James Di Francesco\altaffilmark{2,3},
Sean M. Andrews\altaffilmark{4},
A. Meredith Hughes\altaffilmark{5},
Luca Ricci\altaffilmark{4},
John Bally\altaffilmark{6},
Doug Johnstone\altaffilmark{2,3,7},
and
Brenda Matthews\altaffilmark{2,3}
}
\altaffiltext{1}{Institute for Astronomy, University of Hawaii, Honolulu, HI 96816, USA; jpw@ifa.hawaii.edu}
\altaffiltext{2}{NRC Herzberg Astronomy and Astrophysics, 5071 West Saanich Road, Victoria, BC, V9E 2E7, Canada}
\altaffiltext{3}{Department of Physics and Astronomy, University of Victoria, Victoria, BC, V8P 1A1, Canada}
\altaffiltext{4}{Harvard-Smithsonian Center for Astrophysics, 60 Garden Street, Cambridge, MA 02138, USA}
\altaffiltext{5}{Van Vleck Observatory, Astronomy Department, Wesleyan University, 96 Foss Hill Drive, Middletown, CT 06459, USA}
\altaffiltext{6}{CASA, University of Colorado, CB 389, Boulder, CO 80309, USA}
\altaffiltext{7}{Joint Astronomy Centre, 660 North A’ohoku Place, University Park, Hilo, HI 96720, USA}

\begin{abstract}
We present \emph{ALMA} observations of a wide binary system in Orion,
with projected separation 440\,AU, in which we detect submillimeter
emission from the protoplanetary disks around each star.
Both disks appear moderately massive and have strong line emission
in CO 3--2, HCO$^+$ 4--3, and HCN 3--2.
In addition, CS 7--6 is detected in one disk.
The line-to-continuum ratios are similar for the two disks in
each of the lines.
From the resolved velocity gradients across each disk,
we constrain the masses of the central stars,
and show consistency with optical-infrared spectroscopy,
both indicative of a high mass ratio $\sim 9$.
The small difference between the systemic velocities indicates
that the binary orbital plane is close to face-on.
The angle between the projected disk rotation axes is very high,
$\sim 72^\circ$,
showing that the system did not form from a single massive disk
or a rigidly rotating cloud core.
This finding, which adds to related evidence from disk geometries
in other systems, protostellar outflows, stellar rotation,
and similar recent \emph{ALMA} results, demonstrates that
turbulence or dynamical interactions act on small scales
well below that of molecular cores during the early stages of
star formation.
\end{abstract}


\section{Introduction}
About one in two stars of solar mass and greater are born in
pairs \citep{2013ARA&A..51..269D}.
Because of their common origin and age, young stellar
binaries provide useful benchmarks for understanding
star formation and evolution, and are extensively studied
\citep{1994ARA&A..32..465M}.
Circumstellar disks can exist around the individual stars in
wide systems with semi-major axes $\simgt 100$\,AU, with
masses and lifetimes that are similar to those around single stars
\citep{1996ApJ...458..312J, 2009ApJ...696L..84C, 2012ApJ...745...19K, 2012ApJ...751..115H}
and apparently similar planetary end-products
\citep{2007A&A...462..345D}.

Circumstantial evidence has long suggested that the rotation axes of
such wide binaries are mis-aligned. These include measurements of
stellar rotation \citep{1994AJ....107..306H}
and non-parallel protostellar jets
\citep[e.g.,][]{2002ApJ...576..294L, 2008ApJ...686L.107C}.
More recently, detailed modeling of infrared interferometry
and spectral energy distributions have indicated non-aligned
disk planes in the T Tau \citep{2009A&A...502..623R},
and GV Tau \citep{2011A&A...534A..33R} systems.
With the advent of the
Atacama Large Millimeter/Submillimeter Array (\emph{ALMA}),
the rotation of individual disks
in wide binary systems has now been directly measured and shown
to be misaligned in HK Tau \citep{2014Natur.511..567J}
and AS 205 \citep{2014ApJ...792...68S}.
These observations demonstrate that wide binaries do not form
in large, co-rotating structures and indicate the importance
of stochastic processes during the early phases of star formation,
either gas turbulence or dynamical interactions of young protostars.
Planetary systems that form in such mis-aligned systems may
be subject to secular torques that can affect their orbital
evolution \citep{2012Natur.491..418B}.

The subject of this paper is V2434\,Ori in the M43 HII region of Orion.
\emph{HST} imaging by \citet{2005AJ....129..382S}
reveal this to be a binary system with an angular separation of $1\farcs 1$,
corresponding to 440\,AU at our assumed distance to Orion of 400\,pc
\citep{2007ApJ...667.1161S, 2007A&A...474..515M}.
The optically fainter component is surrounded by a large silhouette disk
and drives a jet with associated Herbig-Haro (HH\,668) objects.
Following the naming convention for \emph{HST}-identified protoplanetary
disks (``proplyds'') proposed by \citet{1994ApJ...436..194O},
we refer to the system as 253-1536.

The discovery that both binary members have disks was made
by \citet{2009ApJ...699L..55M} though submillimeter imaging.
The two disks were subsequently detected at 7\,mm by
\citet{2011ApJ...739L...8R}
and the near-blackbody millimeter colors indicate that both
harbor a substantial population of large dust grains,
characteristic of protoplanetary disks.
Following the nomenclature in those papers, we denote the brighter
millimeter source as component A, although it is the fainter optical source.
Here we present new \emph{ALMA} observations
that reveal the molecular line emission from the disks.
These data allow us to examine disk masses, kinematics, and chemistry,
as well as constrain the masses of their central stars.
Similar to the aforementioned HK Tau and AS 205 results,
we find here that the disks in this system are strongly mis-aligned.
The observations are described in \S2, the results are presented in \S3,
and the implications are discussed in \S4.

\clearpage
\section{Observations}
The data analyzed here come from the fifth field observed in the \emph{ALMA}
Cycle 0 (project 2011.0.00028.S)
study of the Orion proplyds by \citet{2014ApJ...784...82M}
wherein the details of the data acquisition and reduction can be found.
That paper discusses the disk dust masses as inferred from the
$856\,\mu$m (350.5\,GHz) continuum but here we
focus on the molecular lines observed in same observations;
CO 3--2 and CS 7--6 in the lower sideband,
and HCO$^+$ 4--3, and HCN 4--3 in the upper sideband.
The Hanning smoothed spectral resolution of these data is
$d\nu = 488.28$\,kHz corresponding to velocity channels
of $dv \simeq 0.42$\,\kms.

The full extent of the two disks in the binary system is
less than the maximum recoverable scale of $5''$ so
no emission is resolved out.
The resolution of the continuum and line maps, $\sim 0\farcs 5$,
is lower than the Submillimeter Array (\emph{SMA}) images of
\citet{2009ApJ...699L..55M}
but the sensitivity of the \emph{ALMA} data is much higher, allowing us to
detect all four molecular lines and to measure velocity gradients
within each disk.

\section{Results}
\subsection{Moment Maps}
\begin{figure}
\includegraphics[width=3.6in]{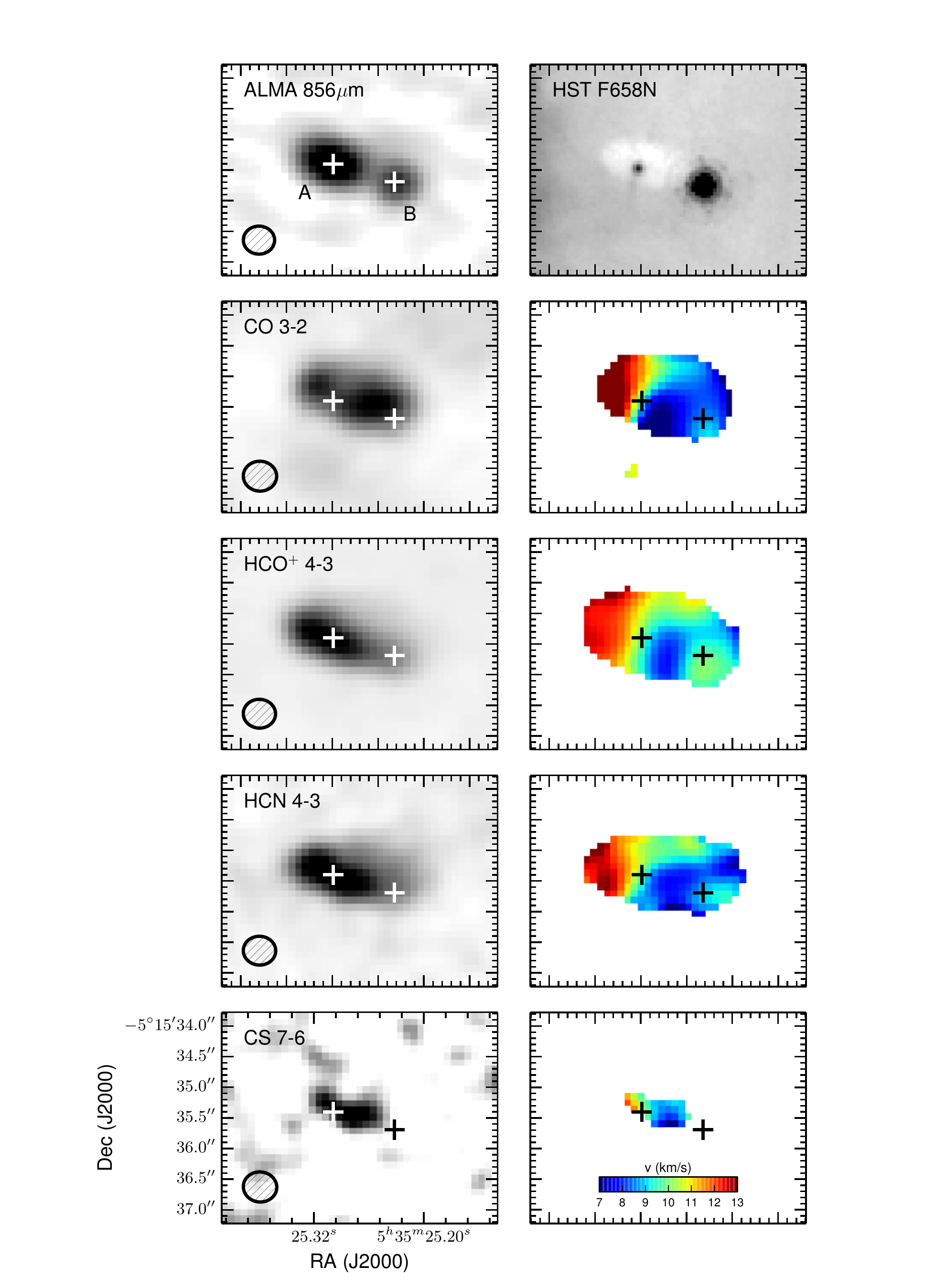}
\caption{
Images of the continuum and line data.
The top row shows the \emph{ALMA} continuum emission at the
central observing wavelength of $856\,\mu$m, and the \emph{HST}
image in the F658 narrow band (H$\alpha$) filter.
The \emph{HST} image shows the two stars and the large silhouette
disk around the fainter source, labeled component A on account of
its brighter millimeter emission. A faint optical jet is seen
perpendicular to the disk, and a diffraction pattern is seen around 
the optically brighter component B.
The \emph{ALMA} image shows disk dust emission from both binary members.
The lower four rows show velocity moment 0 and 1 maps for the
four observed lines. Line emission from CO 3--2, HCO$^+$ 4--3,
and HCN 4--3 is detected toward both disks, and CS 7--6 toward
the large silhouette disk around component A.
The velocity maps are on the same scale, 7 to 13\,\kms,
and show a similar pattern of gradients across each source
but oriented in very different directions.
}
\label{fig:momentmaps}
\end{figure}

Maps of the submillimeter continuum, optical \emph{HST} image,
and molecular lines are displayed in Figure~\ref{fig:momentmaps}.
Both disks are clearly detected in the continuum,
CO 3--2, HCO$^+$ 4--3, and HCN 4--3.
The large silhouette disk around 253-1536A, the brighter millimeter
source, is also detected in CS 7--6.
The positions of the two continuum peaks are given
in Table~\ref{tab:positions}.

The mapped area (primary beam) is much larger than the region shown in
Figure~\ref{fig:momentmaps}. From inspection of the full maps,
we see strong CO emission over a range of size scales from the
background molecular cloud but much weaker emission in the other lines.
The CO contamination affects the disk morphologies,
which show a small offset between line and continuum peaks,
and the resulting flux measurements.
However, the contamination in the other lines is negligible.
The uncontaminated HCO$^+$ line is in fact the strongest line
in the bandpass.  The HCN line is $\sim 4-5$ times weaker,
and the CS line $>25$ times weaker.

The continuum and integrated line fluxes for each disk are
given in Table~\ref{tab:fluxes}.
The error due to flux calibration is estimated to be about 10\%
\citep{2014ApJ...784...82M}.
The continuum flux density is stronger than that measured by
the \emph{SMA} measurements by
\citet{2009ApJ...699L..55M}, partly due to the slightly
shorter observing wavelength but mostly to the greater
quality (phase coherence and signal-to-noise ratio) of the \emph{ALMA} data.
The CO fluxes are unreliable measures of the true disk emission due to
the aforementioned cloud confusion.
The main uncertainty in the fluxes of the other lines,
estimated to be about 20\%, is in separating the overlap between the
two disks as the detectable gas emission extends further than the dust
\citep[e.g.,][]{2008ApJ...678.1119H, 2012ApJ...744..162A}.

\begin{deluxetable}{lcc}
\tablecolumns{3}
\tablewidth{0pt}
\tablecaption{Continuum source positions\label{tab:positions}}
\tablehead{
\colhead{Source} & \colhead{R.A.} & \colhead{Dec.}
}
\startdata
253-1536A & 05:35:25.30 & -05:15:35.40  \\
253-1536B & 05:35:25.23 & -05:15:35.69  \\[-3mm]
\enddata
\end{deluxetable}

\begin{deluxetable}{lccl}
\tablecolumns{4}
\tablewidth{0pt}
\tablecaption{Integrated Flux Measurements\label{tab:fluxes}}
\tablehead{
\colhead{Line} & \colhead{253-1536A} & \colhead{253-1536B} & \colhead{Units}
}
\startdata
Continuum\tablenotemark{a}    &  0.163 &   0.061 & Jy \\
CO 3--2\tablenotemark{b}      &  5.9   &   2.1   & Jy km\,s$^{-1}$ \\
HCO$^+$ 4--3\tablenotemark{c} & 10.5   &   2.7   & Jy km\,s$^{-1}$ \\
HCN 4--3\tablenotemark{c}     &  2.6   &   0.6   & Jy km\,s$^{-1}$ \\
CS 7--6\tablenotemark{c}      &  0.2   & $<0.1$  & Jy km\,s$^{-1}$ \\[-3mm]
\enddata
\tablenotetext{a}{Uncertainty due to flux calibration $\sim 10\%$.}
\tablenotetext{b}{Unreliable measures of disk emission due to strong cloud contamination.}
\tablenotetext{c}{Uncertainty $\sim 30\%$ due to flux calibration
and source overlap.}
\end{deluxetable}

The first moment (intensity weighted mean velocity) of each line
is shown in the right panels of Figure~\ref{fig:momentmaps}
and reveal consistent velocity gradients across both disks.
The direction of the gradients in the two disks are very different
from each other, which indicates that they do not share
the same axis of rotation.
We use the HCO$^+$ data to analyze the velocity structure in more
detail as they provide the best combination of high signal-to-noise
ratio and low cloud confusion.

\subsection{Analysis of HCO$^+$ data}
\begin{figure}[b]
\includegraphics[width=3.3in]{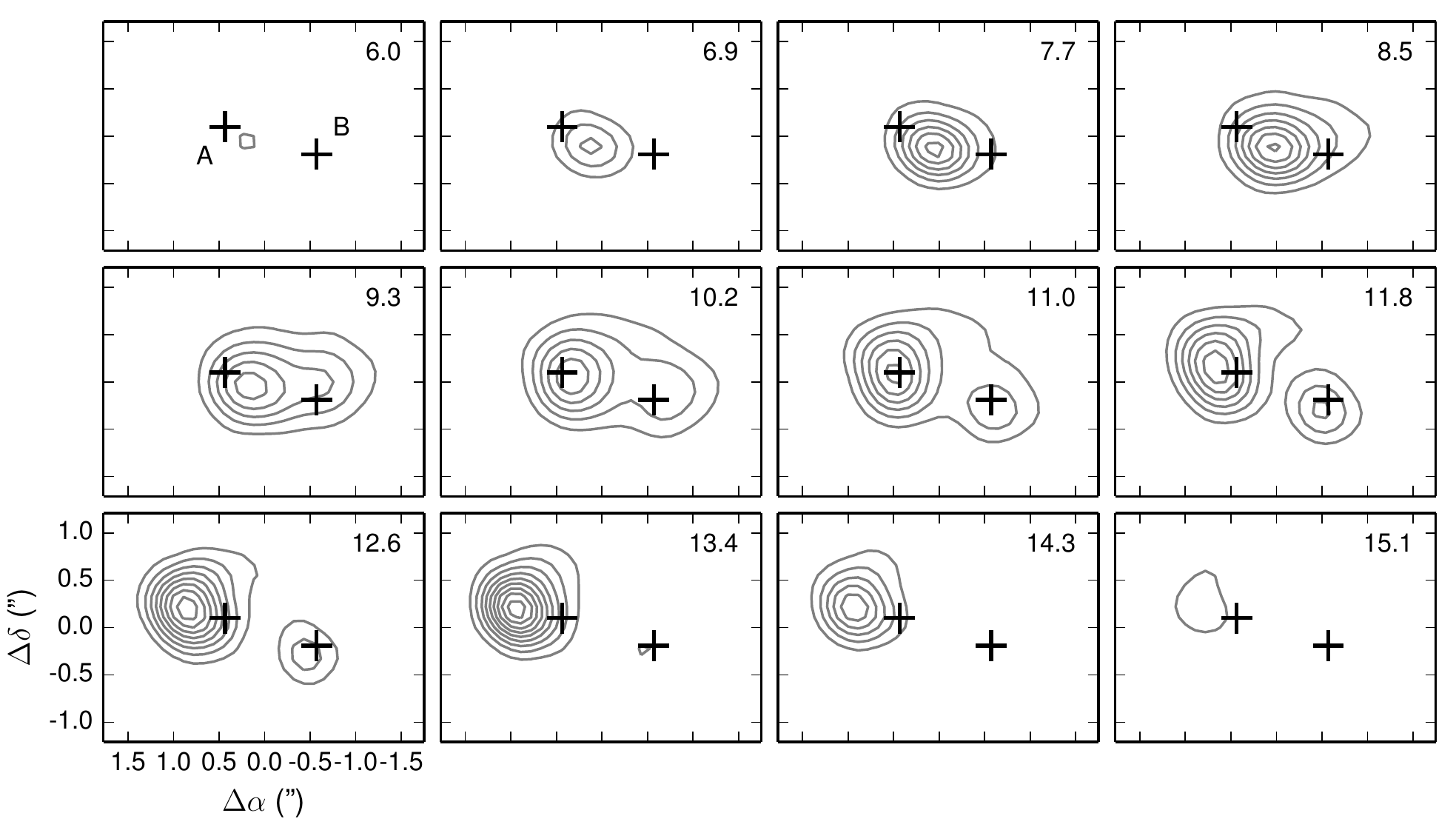}
\caption{
Contours of HCO$^+$ 4--3 emission in velocity channels from 6 to 15\,\kms.
The contour levels start at and increase by 0.1\,Jy\,\kms.
The source central positions are indicated by crosses 
and labeled in the upper left panel.
The velocity of each channel (in \kms) is given in the upper right
corner of each panel.
}
\label{fig:chanmaps}
\end{figure}

Channel maps of the HCO$^+$ 4--3 emission are plotted
in Figure~\ref{fig:chanmaps}. These more clearly show the
velocity gradients across each of the two sources
and the difference in the angle between them.

\begin{figure}[h]
\includegraphics[width=3.4in]{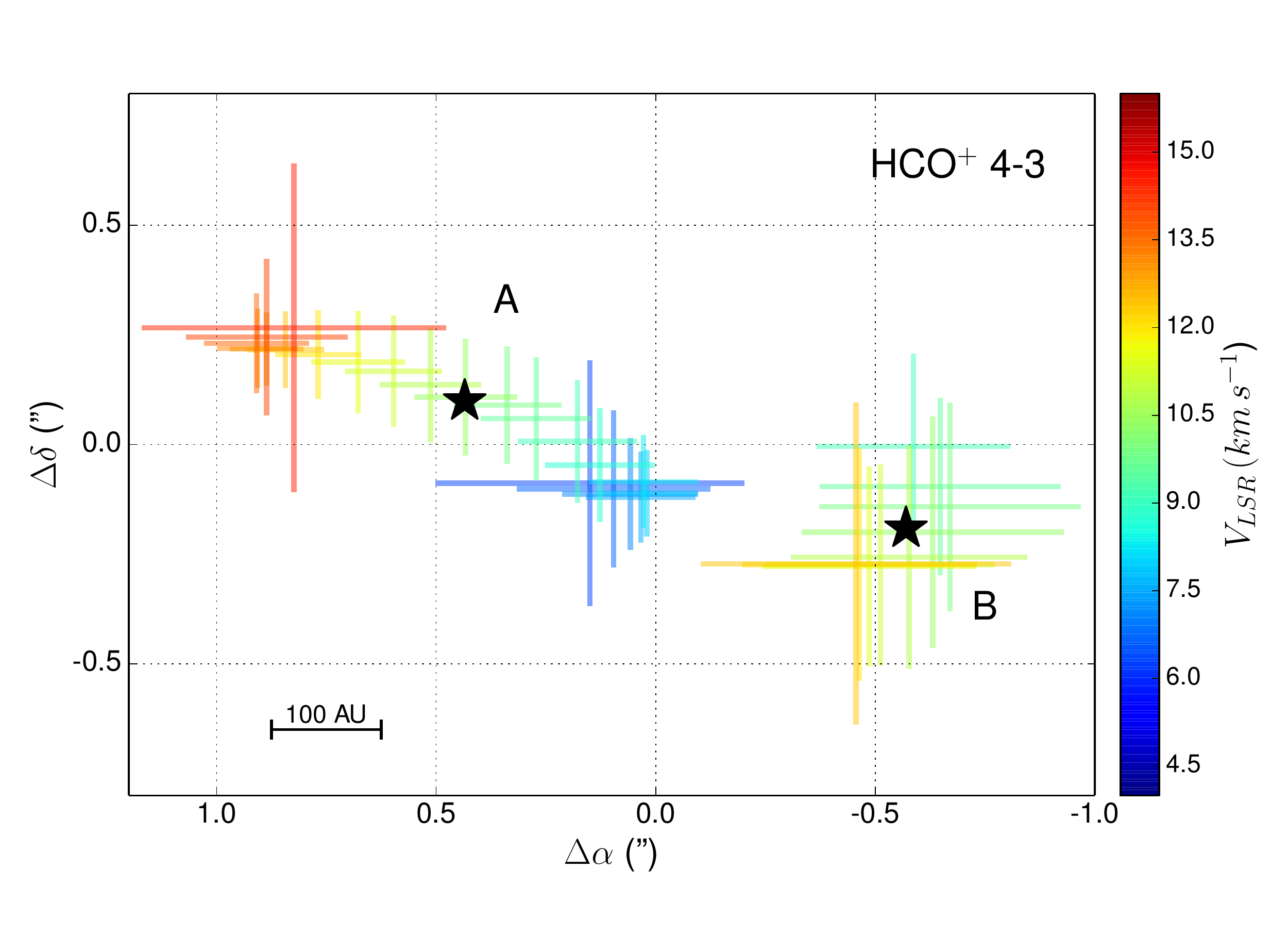}
\caption{
The centroids of each velocity channel in the HCO$^+$ 4--3 datacube.
The cross displays the center and uncertainty associated
with an elliptical gaussian fit to each channel, and the color indicates
the velocity as shown by the scale on the right hand side.
There are clear gradients in the disks around both sources,
and the projected rotational angles are almost perpendicular to each
other. The highest velocity channels for component A are closer to the
star than slightly lower velocities indicative of gravitational motion.
}
\label{fig:centroid}
\end{figure}

To measure the size and direction of the velocity gradient, we fit
elliptical gaussians to the channel maps
as in \citet{2012Natur.492...83T}.
The centroids of the fits toward the two sources are shown,
color-coded by velocity, in Figure~\ref{fig:centroid}.
The reversal of the
centroids at the lowest and highest velocities toward the
large silhouette disk around 253-1536A is a clear signature of
gravitational motion; the gas moves faster closer to the star.
Because of the weaker emission toward 253-1536B, fewer channels
were detected and we are unable to see the same signature.
Given the strength and small velocity extent of the HCO$^+$
and HCN lines, however, we assume that they similarly trace
the kinematics of the dusty disk rather than an outflow.

\begin{figure}
\includegraphics[width=3.5in]{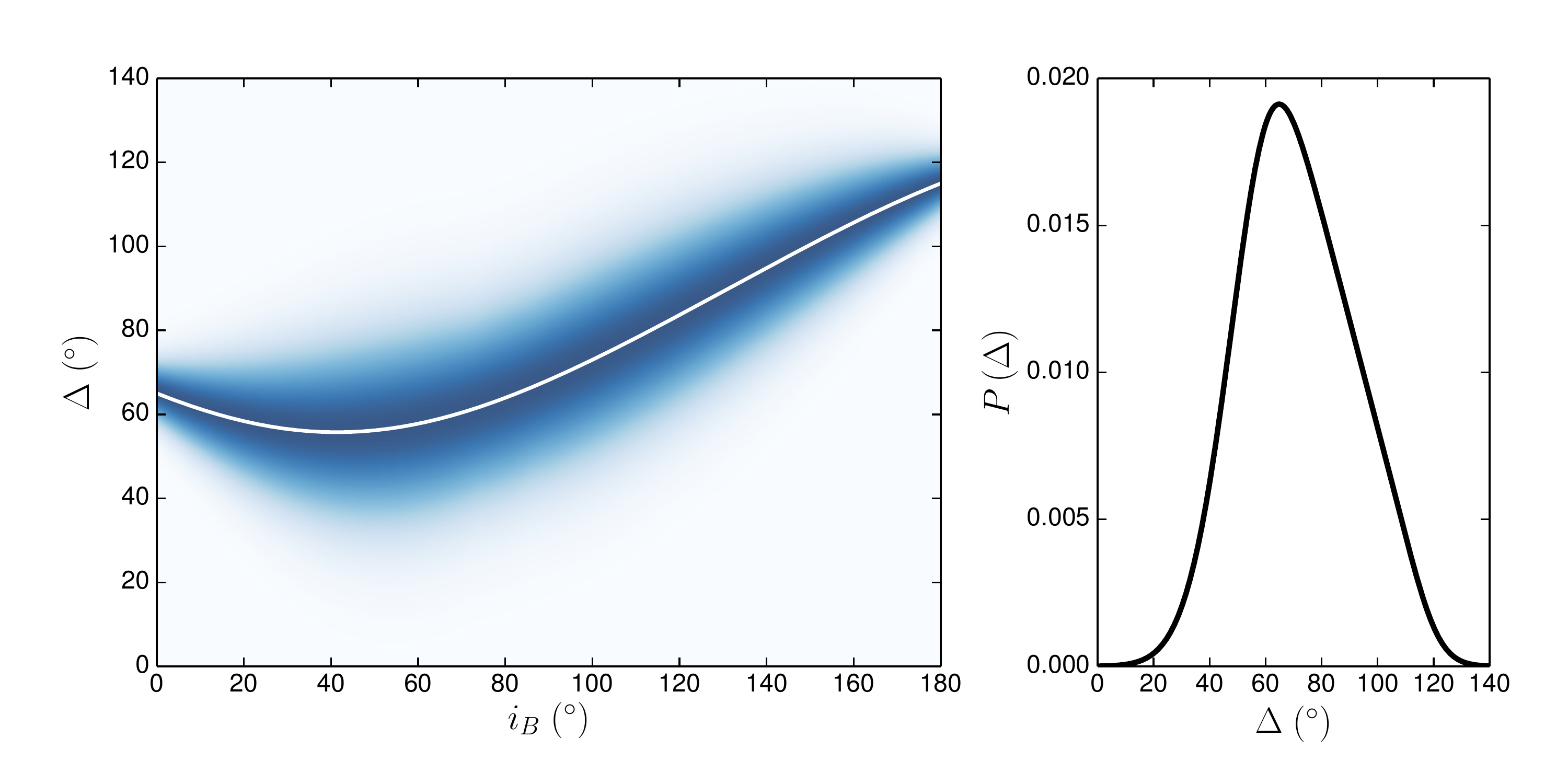}
\caption{
The left panel shows the angle, $\Delta$, between the disk rotational
axes, as a function of the unknown inclination of the unresolved
source 253-1536B, $i_{\rm B}$.
The (assumed gaussian) uncertainty at each value is illustrated
by blue shading, and is due to the errors in the measured
inclination of source A and projected position angles of the two disks.
The right panel shows the posterior probability (per degree) of $\Delta$
given a prior for the inclination, $P(i_{\rm B})\propto\sin(i_{\rm B})$.
}
\label{fig:angle}
\end{figure}

Through linear fits to the positions of the centroids,
we determine the projected rotational planes of the two disks.
Following the convention that the position angle (PA)
is measured east of north to the redshifted edge of the disk
we find $PA_{\rm A}=69.7\pm 1.4^\circ$ and $PA_{\rm B}=136\pm 15^\circ$
for components A and B respectively\footnote{
As the data are Hanning smoothed, we binned the velocity channels
by 2 to provide independent points for the purpose of making the linear fits.
}.
These are projections and the true angle between the rotational
axes of the disks
(derived from the dot product of the angular momentum vectors)
depends on the inclinations of the two disks,
\begin{equation}
\cos\Delta = \cos i_{\rm A}\,\cos i_{\rm B}
           + \sin i_{\rm A}\,\sin i_{\rm B}\,\cos(PA_{\rm A}-PA_{\rm B}),
\end{equation}
as in \citet{2014Natur.511..567J}.
For the resolved silhouette disk, 253-1536A, we determine
an inclination from face-on, $i_{\rm A}=65\pm 5^\circ$,
based on the $0\farcs 6\times 1\farcs 4$ size of the optical
shadow \citep{2005AJ....129..382S}, but the inclination of
the unresolved disk, 253-1536B, is unknown and can vary from
$0^\circ$ to $180^\circ$ where $90^\circ$ is edge-on.
We plot the angle, $\Delta$, as a function of this inclination
in the left panel of Figure~\ref{fig:angle}.
For random orientations, the probability of a particular
inclination is proportional to the sine of the inclination
(i.e., edge-on disks are more common than face-on).
Using this as a prior, and without attempting to fit the
observations further, we plot the posterior probability distribution
for $\Delta$ in the right panel of Figure~\ref{fig:angle}
and derive a mean value and standard deviation,
$\Delta = 72^\circ\pm 20^\circ$.
Thus we can robustly conclude that the disks are indeed
strongly mis-aligned.

\begin{figure}
\includegraphics[width=3.8in]{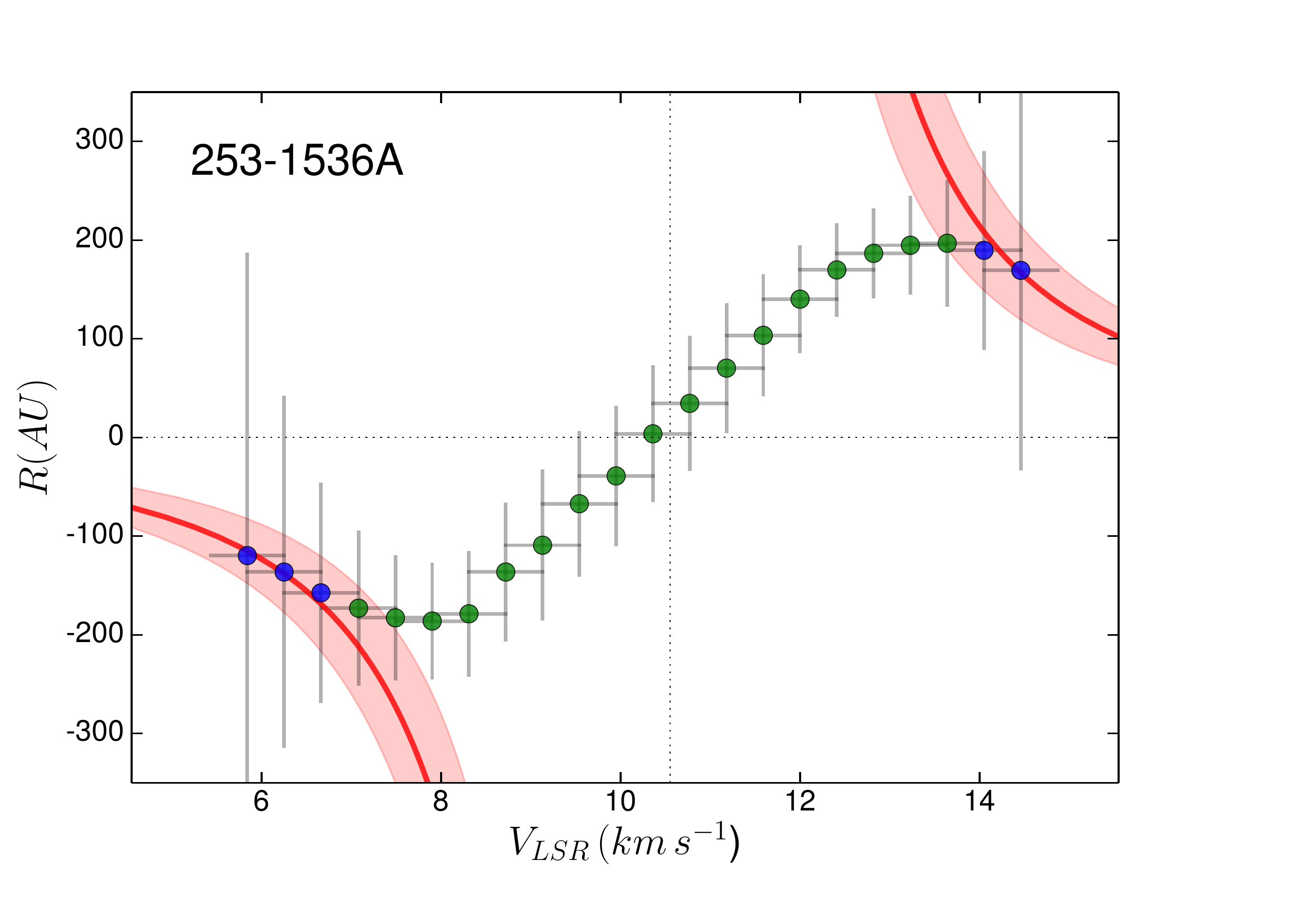}
\caption{
The distance from the star in binary component A
for each velocity channel of the disk HCO$^+$ emission.
The reversal in distance at the greatest speeds away from systemic are
shown as blue points and compared with Keplerian profiles for a star
with mass $M_{\rm star}=3.5\,M_\odot$ (solid red line) and range
$\pm 1\,M_\odot$ (light red shaded region).
}
\label{fig:rotation}
\end{figure}

The projected radius from the star is plotted against the channel
velocity for the silhouette disk around 253-1536A in
Figure~\ref{fig:rotation}.
The S-shape is due to a combination of resolution, sensitivity,
and Keplerian rotation. Because the disk is inclined and not resolved
along the minor axis, emission at velocities close to the systemic
motion of the star is dominated by the bright emission from the inner
regions of the disk that project to low radial velocities
(rather than the fainter emission from the slower moving outer disk)
and channel maps only show slight offsets with respect to the star.
There is less confusion at greater relative velocities but also
weaker emission. There are a few low and high velocity channels,
however, for which we can detect the increasing rotational
velocity closer to the star. These are marked in blue and compared
with a Keplerian profile,
\begin{equation}
v_{\rm LSR} = v_{\rm sys} + \left(\frac{GM_*}{R}\right)^{1/2}\sin\,i,
\end{equation}
in Figure~\ref{fig:rotation}
to provide a rough estimate of the central stellar mass.
The resolution and signal-to-noise ratio of the data are
insufficient to attempt a more detailed model and more rigorous
fitting as in e.g., \citet{2012ApJ...759..119R}.
With a well constrained inclination, $i_{\rm A}=65^\circ$,
as discussed above, we infer $M_*\sim 3.5\,M_\odot$.

The moderately high stellar mass for this optically faint star is
consistent with the X-Shooter ultraviolet-optical-infrared spectrum
discussed by \citet{2011ApJ...739L...8R}.
The lack of prominent absorption lines led them to conclude
that the star is heavily veiled and is spectral type F or G.
For an age range of 1--3\,Myr, this corresponds to a mass,
$M_*\sim 2.5-4\,M_\odot$.

We can similarly look at the rotation curve for 253-1536B.
The disk emission is weaker, both in the line and continuum,
and we do not resolve it spatially.
However, we are able to measure a shift in the peak position
in different spectral channels and from this can study its
kinematics. Although we do not detect a Keplerian turnover,
the velocity gradient in Figure~\ref{fig:centroid}
provides a lower limit to the stellar mass,
$M_* \gtrsim 0.2\,M_\odot/\sin^2\,i_{\rm B}$.
This is consistent with the spectral typing of an M2 star
($M_*\sim 0.4\,M_\odot$ for an age range 1-3\,Myr)
from the X-Shooter spectrum \citep{2011ApJ...739L...8R}
and the catalog of \citet{1997AJ....113.1733H},
which classifies it as M2.5e (source 767).
Both sets of authors note emission lines in the spectrum
that are signatures of strong accretion.

The central velocities of 253-1536A and B are $v_{\rm sys}=10.55$
and 10.85\,\kms\ respectively, with errors of about 0.1\,\kms.
The escape speed of the two stars from one another is
$\sim 2.5$\,\kms\ at the projected separation of 440\,AU,
and greater than the measured 0.3\,\kms\ difference unless
the system is $\simgt 10^4$\,AU apart and viewed almost edge-on.
However, the probability of a chance alignment of the two stars
was already known to be low \citep[see discussion in][]{2009ApJ...699L..55M}
and this additional kinematic information more likely indicates that
this is a bound binary system with an orbital plane close to face-on.

\section{Summary and Discussion}
\label{sec:discussion}
The tremendous increase in sensitivity that \emph{ALMA} provides
is transforming our view of protoplanetary disks.
Whereas we had known of the existence of the two disks
under study here from \emph{SMA} continuum observations,
we can now detect their line emission and study their kinematics.
The magnitude of the measured velocity gradients provides bounds
on the central masses that are consistent with stellar spectroscopy.
The system has a very high mass ratio, $\sim 9$, that makes it a
useful laboratory for studying the dependence of disk properties
on stellar mass.

From the continuum alone, we estimate disk masses of
$0.074\,M_\odot$ and $0.028\,M_\odot$ for sources 253-1536A and B
respectively, assuming canonical values for the temperature, dust opacity
and an ISM gas-to-dust ratio of 100 \citep{2011ARA&A..49...67W}.
Under these assumptions, the disk to stellar mass ratios are
$\sim 2\%$ and $\sim 7\%$ respectively.
This is particularly high for for the M2 star, 253-1536B,
but the dust is likely warmer than the canonical 20\,K
(possibly in both disks) given the high stellar mass and
therefore luminosity of component A.
Following the expected $L^{1/4}$ dependence from
\citet{2013ApJ...771..129A}, we might expect a factor of
2--3 increase in average dust temperature and a correspondingly
decrease in the dust masses.
These temperatures, masses, and mass ratios can only be estimates
given the limited information but serve as a comparison with other
millimeter wavelength disk observations.

The strong emission in multiple molecular lines suggests significant
amounts of moderately warm and dense gas in both disks.
The integrated line fluxes are only about a factor of 3--5 weaker in
253-1536B compared to the brighter millimeter component A,
despite its much lower stellar luminosity.
Radiation from the nearby primary member may be a significant
factor in heating both disks and could potentially explain
their roughly similar (to within $\sim 50$\%) line-to-continuum ratios.
To compare the chemistry in the two disks, and also to measure
their gas masses and gas-to-dust ratios, requires observations of
optically thinner isotopologues \citep{2014ApJ...788...59W}.

Perhaps the most noteworthy result from these observations are that
the projected disk rotational axes are highly misaligned,
with an angle of $72\pm 20^\circ$ to each other.
As torques from the binary orbit act to align the disks
on relatively short timescales \citep{2000ApJ...538..326L},
the observed conditions are most likely a signature of their formation.
Similar \emph{ALMA} kinematic measurements of mis-alignment
have been recently found in two other binary disk systems
\citep{2014Natur.511..567J, 2014ApJ...792...68S}.

These observations demonstrate that wide binaries do not form from
the same co-rotating structure such as a massive disk or coherent
cloud core \citep{2000prpl.conf..675B}.
Signatures of disorder from the early phases of star
formation persist. One possibility is the fragmentation of a turbulent
core \citep{2010ApJ...725.1485O, 2014ApJ...789L...4T}.
Numerical simulations shows that binary formation in even weakly
turbulent cloud cores is quantitatively different than the purely
thermal case \citep{2013MNRAS.428.1321T}.
In turbulent cores, the direction of angular momentum vectors
vary with spatial scale such that disks may form at different
angles from each other and from the overall core rotation axis
\citep{2012MNRAS.419.3115B}.

Alternatively, dynamical interactions of three or more protostars during
the early Class O-I phases may chaotically scramble orbital axes.
Non-hierarchical triple or small multiple systems with similar
interstellar separations rapidly rearrange into hierarchical configurations
consisting of a compact binary and distant companions or ejected members
\citep{2012Natur.492..221R, 2010ApJ...725L..56R}.
The observed decrease in multiplicity from Class 0 to I to main-sequence
stars provides support for such dynamical evolution
\citep{2013ApJ...768..110C}.
In crowded regions, large disks may also assist in the capture of
binaries, such as proposed for the massive star system Cepheus A
\citep{2009ApJ...692..943C}.
A similar event is thought to have occurred within the last 500 years
in Orion BN/KL
\citep{2011ApJ...727..113B, 2011ApJ...728...15G}.

Stellar rotation and orbital axes are more tightly aligned
in closer systems, $a\simlt 40$\,AU
\citep{1994AJ....107..306H},
than those studied with \emph{ALMA} to date.
This is also seen in numerical simulations of cluster formation
\citep{2012MNRAS.419.3115B}.
It would be interesting to search for disk alignment
in such close binaries but,
because the same proximity that might align the disks also tidally
truncates them \citep{1994ApJ...421..651A, 2010ApJ...710..462A}
and lowers their masses \citep{2009ApJ...696L..84C},
higher resolution and signal-to-noise observations
than shown here will be required.

At larger scales, one can imagine larger studies across star forming
complexes providing new information on the velocity dispersion
of protostars in different evolutionary states that can constrain
timescales and the turbulent properties of the cloud from which
they formed.
As disk surveys continue in new \emph{ALMA} observing cycles,
including an expanded program of Orion proplyds,
we can expect to routinely detect a much broader suite
of molecules and transitions than has been observed
in all but a handful of the brightest disks to date.
This will open up new paths of exploration for not only studying
statistical disk properties such as masses, sizes, gas-to-dust ratios
and chemistry, but also for examining the dynamics of young star
forming regions.

\acknowledgments
We thank the referee for a very thorough report,
Kaitlin Kratter, Stella Offner, and Hideko Nomura for comments,
and Eric Jensen and Rachel Akeson for communicating their results
ahead of publication.
J.P.W. is supported by funding from the NSF through grant AST-1208911.
D. J. is supported by the National Research Council of Canada and by a
Natural Sciences and Engineering Research Council of Canada (NSERC)
Discovery Grant. 
ALMA is a partnership of ESO (representing its member states), NSF (USA)
and NINS (Japan), together with NRC (Canada) and NSC and ASIAA (Taiwan),
in cooperation with the Republic of Chile.
This work made use of Astropy, a community-developed core Python package for
Astronomy \citep{2013A&A...558A..33A}.

{\it Facilities:} \facility{ALMA}.


\end{document}